\documentclass[aps,amssymb,showpacs,superscriptaddress]{revtex4}
\usepackage{amsmath} \usepackage{amssymb} \usepackage{amsfonts}
\usepackage{graphicx}
\usepackage{times}

\usepackage{textcomp} \usepackage[normalem]{ulem}
\usepackage[usenames]{color}

\begin{document} \title{Dynamics of an impurity in a one-dimensional
lattice} \author{F. Massel} \altaffiliation[Present address: ]{Department of Mathematics and Statistics, University of Helsinki, P. O. Box 68 Fin-00014, Helsinki, Finland} \affiliation{Olli V. Lounasmaa Laboratory,
Aalto University, FI-00076 Aalto, Finland} \author{A. Kantian}
\affiliation{DPMC-MaNEP, University of Geneva, 24 Quai Ernest
Ansermet, 1211 Geneva, Switzerland} \author{A. J. Daley} \affiliation{
Department of Physics and Astronomy, University of Pittsburgh,
Pittsburgh, Pennsylvania 15213, USA} \author{T. Giamarchi}
\affiliation{DPMC-MaNEP, University of Geneva, 24 Quai Ernest
Ansermet, 1211 Geneva, Switzerland} \author{P. T\"orm\"a}
\affiliation{ COMP Centre of Excellence, Department of Applied
Physics, Aalto University, FI-00076 Aalto, Finland}

\begin{abstract} We study the non-equilibrium dynamics of an impurity
in an harmonic trap that is kicked with a well-defined quasi-momentum,
and interacts with a bath of free fermions or interacting bosons in a
1D lattice configuration. Using numerical and analytical techniques we
investigate the full dynamics beyond linear response, which allows us
to quantitatively characterise states of the impurity in the bath for
different parameter regimes. These vary from a tightly bound molecular
state in a strongly interacting limit to a polaron (dressed impurity)
and a free particle for weak interactions, with composite behaviour in
the intermediate regime. These dynamics and different parameter
regimes should be readily realizable in systems of cold atoms in
optical lattices.
\end{abstract} 
\maketitle

\pagebreak Impurities play a crucial role in determining the
low-temperature features of a number of condensed matter systems.
These impurities may be localized ones, as for the x-ray edge
\cite{Mahan:2010wd} and Kondo effects \cite{Kondo:1964eg}, or mobile
ones, like the itinerant single electrons modified by the phonon bath
of the solid state crystal in which they move (polarons)
\cite{Mahan:2010wd}, or the single spin-flipped electron moving in a
lattice populated by opposite spin electrons, as studied in the
context of high $T_c$ superconductors \cite{Lee:2006de}. With recent
experimental developments for systems of ultracold atoms, such as the
the tunability of the two-body interaction with the aid of Feshbach
resonances \cite{Ketterle:1998if}, a particularly well controlled
environment \cite{Bloch:2008gl} to explore the properties of these
types of many-body systems has become available.

For two- and three-dimensional systems, these advances have enabled
the experimental \cite{Schirotzek:2009jw,
Nascimbene:2009cu,Sommer:2011ig,Koschorreck:2012fj,Kohstall:2012kg,Zhang:2012dj}
and theoretical \cite{
Chevy:2006hv,Prokofev:2008jz,Massignan:2008hn,Punk:2009gd,Zollner:2011cc,Parish:2011eo,Massignan:2011ey,
Giraud:2012gx,Schmidt:2012cd,Baarsma:2012if,Ngampruetikorn:2012ge,Knap:2012vf}
study of mobile impurities inside a fermionic bath, i.e. a type of
polaron, usually created by preparing two-component ultracold Fermi
gases with a large number imbalance between the components. In that
context, the possibility of tuning the bath-impurity interaction
across a wide range and even from attractive to repulsive regimes has
opened up both the polaronic and the molecular regime to
investigation. These advances have also stirred interest in using
these fermionic systems to study the dynamics of the x-ray edge
effect, which is induced by a localized impurity \cite{Knap:2012vf}.

At the same time, the ability to restrict the spatial dimension of the
ultracold gas experiments almost arbitrarily has also made the study
of impurities inside one-dimensional many-body baths possible. As
movement of the impurity in such a bath can very easily involve the
collective motion of many bath atoms, the result can be profoundly
modified compared with what would be expected in higher-dimensional
systems, giving rise to a regime of subdiffusive impurity motion, in
which it can displace only proportional to the logarithm of time,
slower than any power law
\cite{Zvonarev:2007,Zvonarev:2009pl,Zvonarev:2009pb,Imambekov:2008,Lamacraft:2009bp}.
Another reason for the particular interest in 1D impurity-bath systems
is that they make particularly compelling benchmark systems for a wide
range of impurity-bath systems, due to the powerful theoretical
approaches available to treat interacting 1D systems
\cite{Giamarchi:2003,Essler:2005uw}. For example, the ground state of
an impurity in a 1D Fermi gas in a lattice was calculated via exact
numerical methods~\cite{Leskinen:2010tm}, demonstrating that it can be
described by a polaron-type ansatz for weak interactions, while the
strong interaction regime corresponds well to the strongly interacting
limit of the Bethe ansatz. Static properties of polarons in 1D
ultracold Fermi gases have been studied also in~\cite{Guan:2012wm},
and recently there has been interest in exploring the dynamics as well
\cite{Mathy:2012tg,Schecter:2012jb}.  Complementary to the fermionic
case, the dynamics of an impurity in a continuous bosonic bath was
studied recently experimentally and
theoretically~\cite{Palzer:2009tc,Catani:2012,Peotta:2012uw,Bonart:2012}.
Major advances with single-site addressing and manipulation in optical
lattices \cite{Bakr:2010gd,Sherson:2010hg} have recently enabled the
realization of lattice impurities within a bath described by a 1D
Bose-Hubbard model \cite{Fukuhara:2012vh}.

In this article we explore the basic dynamical properties of a single
impurity in a lattice potential and a harmonic trap in 1D, which
interacts with a bath of free fermions or interacting bosons, also
confined in the 1D lattice. Specifically, we consider the
non-equilibrium response of the impurity to a kick with well defined
momentum. A key open question in this context is how to characterise
the role that the bath atoms play in the dynamics. In particular, we
ask whether the dynamical response of the system implies polaronic
behaviour, in which the properties of the particle are renormalised by
the presence of the bath, or whether the interaction gives rise to
other states, e.g., to tightly bounds pair or more complex objects.

Using time-evolving block decimation (TEBD) methods
\cite{Vidal:2004,Daley:2004hk,White:2004,Verstraete:2008} in
conjunction with Bethe-ansatz results, we study the non-equilibrium
dynamics of the impurity beyond the weak coupling assumptions of
linear response theory. We show that the observed oscillation
frequencies of the impurity-bath system can be mapped onto different
physical states, and explain their dependency on bath density and
strength of the impurity-bath interaction. In different limits we see
that the behaviour ranges from a tightly bound pair for strong
interactions to polaron-like behaviour at weak interactions for a
fermionic bath. The latter case is characterised by an interesting
internal dynamics corresponding to the scattering between a bound pair
and an impurity particle propagating through the fermionic bath. We
also compare these results with the case of a bosonic bath to better
define the role of the Fermi sea. Generally, we find that the physics
for a boson bath can be qualitatively and even quantitatively similar
to the fermionic case, for both the doublon and the polaron regime,
provided the boson-boson repulsion is larger than the attractive
interaction between bath and impurity.

This setup and characterisation of the dynamics should be readily
realisable with cold atoms in optical lattices, and we expect our
zero-temperature results to hold also at finite temperature, provided
it is lower than the energy scale given by the oscillation frequency.

This article is organised as follows: we first introduce the system
and describe the method used in Section \ref{fermions}. There we also
discuss the effects of combining a lattice potential with a trap in
the case of an impurity that does not interact with the bath. In
Section \ref{sec:pairs-dynamics} we present numerical results for the
impurity dynamics and explain them both in the regime of strong and
weak interactions through comparison with analytical methods. We
especially discuss the frequency spectrum of oscillations, and
identify different physical regimes of impurity behaviour. In Section
\ref{bosons}, we compare with the case of a bosonic reservoir, and
identify similarities and differences to the case of the fermionic
bath dependent on the boson-boson repulsion. In Section
\ref{sec:disc_conc}, we discuss our findings and make a connection to
earlier impurity and polaron studies. Finally, an appendix contains
details of several analytical results we have derived.

\section{Fermionic system}
\label{fermions}

\subsection{Basic model and method}
\label{sec:setup_fermions}

We consider a setup that is constituted by a optical lattice, loaded
with a number-imbalanced mixture of two hyperfine species of fermionic
atoms, hereafter labelled $\uparrow$ and $\downarrow$, which are
confined to move along one dimension.  Our interest lies in the case
of extreme imbalance, namely $N_\downarrow=1$ and
$N_\uparrow\in\{1\dots L\}$, where $L$ is the lattice size and
$N_{\uparrow /\downarrow}$ is the total number of atoms for each
species.  In addition to the optical lattice, the $\downarrow$ atomic
impurity experiences a parabolic confining potential.  For atoms in
the lowest Bloch band, the system can be described by the Hubbard
Hamiltonian ($\hbar\equiv 1$)
\begin{equation}
   \label{eq:Hubb} H= -J\sum_{i \, \sigma} c_{i \sigma}^\dagger c_{i+1
\sigma} + h.c. + U \sum_i n_{i\, \uparrow}n_{i\, \downarrow} + V
\sum_i n_{i \,\downarrow} \left(i-\frac{L-1}{2}\right)^2,
\end{equation} where $J$ represents the hopping amplitude between
neighbouring sites, $U<0$ is the on-site (attractive) interaction
energy and $V$ characterises the strength of the parabolic confining
potential for the impurity. Throughout the paper we set $\hbar=1$, and
we choose as the length scale the lattice period $a$. Therefore, all
energies are given in frequency units and all momenta are given in
units of $1/a$.
 
To obtain the ground state of this Hamiltonian and simulate the full
many-body dynamics after the impurity has received a kick with a
defined quasi-momentum, we use a code based on the TEBD algorithm, for
which more details can be found e.g. in
Refs. \cite{Massel:2009hk,Korolyuk:2010cl,Kajala:2011prl,Kajala:2011pra}. In
our simulations we have considered a lattice size of $L=200$ sites
($L=400$ in one case), $N_\downarrow=1$, $N_\uparrow=[1,\,10,\,20
\dots 180,\, 190,\, 200]$, and $|U|/J=0, 0.5,\, 1,\,3,\,5,\,10,\,20$
with particular emphasis on the case $|U|/J=10$. At $t=0^+$, a
quasi-momentum $k$ ($k=0.1 \pi$ unless otherwise stated) is imparted
to the $\downarrow$ impurity.

\begin{figure}%[!ht]
  \includegraphics[angle=-90,width=0.8\textwidth]{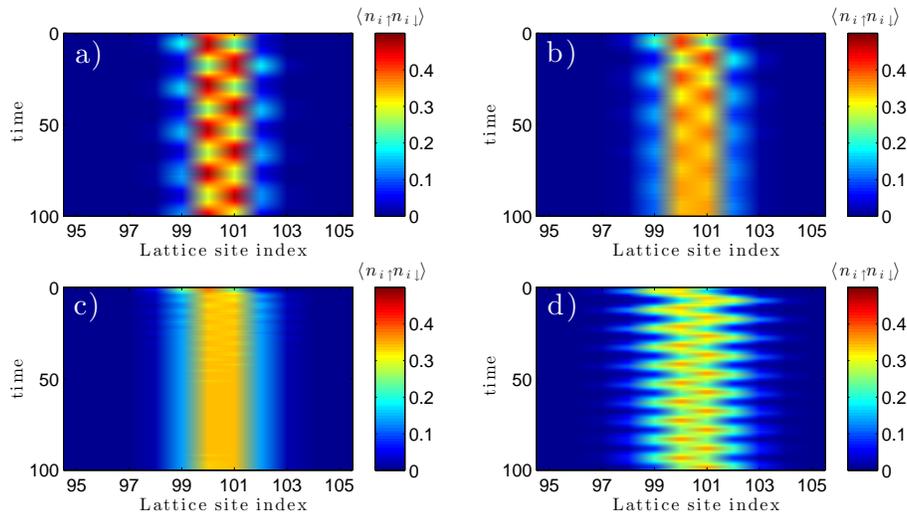}
  \caption{Time dependence of the local doublon density $\langle
n_{i\,\uparrow}n_{i\,\downarrow}\rangle$ for $|U|/J=10$ and
$N_\uparrow$= 20 (a), 60 (b), 140 (c), 180 (d).  Horizontal axis: site
index $i$, vertical axis: time (in units of $J^{-1}$).  In the
simulations we first calculate the ground state of the lattice (number
of lattice sites $L=200$) loaded with one impurity ($N_\downarrow=1$),
in presence of a bath of spin-up particles (different simulations were
performed with varying numbers of up particles
$N_\uparrow=[1,\,10,\,20 \dots 180,\, 190,\, 200]$), whose description
is provided by the Hubbard Hamiltonian. In addition to the attractive
interaction between spin-up and spin-down particles, characterized by
the parameter $U$, the impurity is confined by a parabolic potential
of the form $V i^2$, with $i$ the site index, and $V/J=0.1$. At
$t=0^+$, the down particle is kicked with quasi-momentum $k=0.1
\pi$. As is described in the text, it is possible to see that, as a
function of $N_\uparrow$, the dynamics exhibits different regimes,
characterized by a transition from doublon-oscillations to a regime
where the internal dynamics of the polaron dominates, and eventually
to free-particle oscillations. All simulations are performed with a
TEBD code, with the following numerical parameters: lattice size
$L=200$, unless otherwise stated; Schmidt number $\chi=80$; initial
imaginary timestep (ground-state calculation) $\delta t_i=0.1 J^{-1}$;
timestep (real time evolution) $\delta t=0.02 J^{-1}$ (throughout the
paper $\hbar=1$, and the length scale is set equal to the lattice
spacing $a$, see text). After testing with larger values of the
Schmidt number, we have used the value $\chi=80$ which provides
accurate results both for the ground state and the time evolution. The
reason of the effectiveness of this rather small value of $\chi$ lies
in the reduced size of the Hilbert space for a strongly imbalanced
gas.}
 \label{fig:nudN}
\end{figure}
\begin{figure}%[!ht]
  \includegraphics[angle=-90,width=0.8\textwidth]{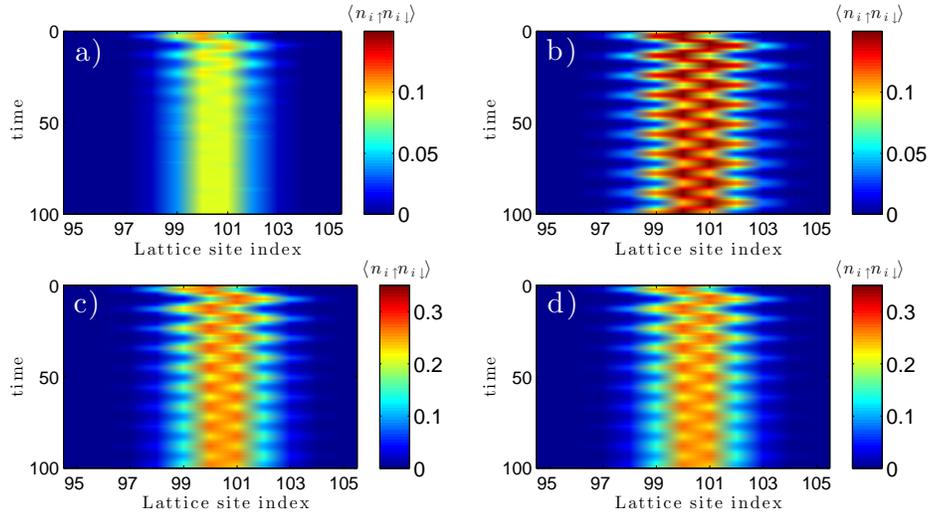}
  \caption{Time dependence of the local doublon density $\langle
n_{i\,\uparrow}n_{i\,\downarrow}\rangle$ for $|U|/J=1$ and
$N_\uparrow$= 20 (a), 60 (b), 140 (c), 180 (d). For $|U|/J=1$, the
time evolution of $\langle n_{i\,\uparrow}n_{i\,\downarrow}\rangle$ is
characterized by single-particle oscillations decreasingly damped for
increasing bath particle population. For $N_\uparrow=20$,
small-amplitude oscillations (not visible here, see
Fig. \ref{fig:cm_w1}) appear; these oscillations are associated with
the pair-breaking mechanism as we will discuss in Sec. \ref{sec:weak}.
All simulation parameters (except $U$) are the same as in
Fig. \ref{fig:nudN}.}
 \label{fig:nudN1}
\end{figure}

\subsection{The non-interacting impurity}
\label{sec:free-particle}

To provide a better description later on of the effects on the motion
of the impurity that are induced by interactions with the bath, we
first single out the dynamical effects induced by the concomitant
presence of the harmonic trapping and the 1D lattice at $U=0$. That
is, we consider a single particle in a potential formed by combining a
lattice and a harmonic potential. As shown in \cite{Rey:2005bw}, if
the particle is initially in the ground state of the combined
potential, and then the harmonic trap is displaced by an amount
$\delta$, the expectation value of the particle position is
\begin{equation}
  \label{eq:COM2} \langle x \rangle = \delta \exp \left[ -\left(
\frac{\delta}{a_{\rm ho}}\right)^2 \sin^2\left(V/8 t\right)\right]
\cos\left[V\left(\sqrt{\frac{2}{V m_{\rm eff}}}-\frac{1}{4}\right)t-
\frac{\delta^2}{2 a_{\rm ho}^2}\sin\left(\frac{V
t}{4}\right)+\frac{\pi}{2}\right] ,
\end{equation} where $V$ is the strength of the trapping potential,
and $m_{\rm eff}$ is the effective mass of the particle in the
lattice, which for our choice of units ( $a=1$) is $m_{\rm eff}=0.5
J^{-1}$. The harmonic oscillator length is defined via the trapping
frequency $\omega$ and the mass of the particle as $a_{\rm
ho}=1/\sqrt{m_{\rm eff} \omega}$, and on the other hand the strength
of the harmonic oscillator potential $V=m_{\rm eff} \omega^2/2$: thus
$a_{\rm ho}=(2 m_{\rm eff} V)^{-1/4}$ and with $m_{\rm eff}=0.5
J^{-1}$ we then should have $a_{\rm ho}=(V/J)^{-1/4}$.

Our initial conditions involving a finite quasi-momentum kick differ
from those in Ref.~\cite{Rey:2005bw}, where the harmonic trap is
initially displaced by an amount $\delta$. To account for this
difference, we need to introduce a $\pi/2$ phase in the cosine term of
Eq.~\ref{eq:COM2}, and also compute the value of $\delta$
corresponding to our initial momentum kick. We can do this by matching
the energy of the initial state in each case. The energy for the state
at $t \to 0^+$ is given by the lowest eigenenergy of the combined
harmonic trap and lattice system (Eq. (14) of \cite{Rey:2005bw}), with
the addition of the kinetic energy given by the kick of a
quasi-momentum $k$. If, by a semiclassical argument, we assume that
this energy must be converted to potential energy (removing the
zero-point energy from both terms), then we have
\begin{equation}
  \label{eq:en_delta} V \left(\delta^2 + \frac{1}{16} + \frac{
\sqrt{V/J}}{256}\right)= (1- \cos k )/m_{\rm eff} .
\end{equation} The left hand side of Eq. \eqref{eq:en_delta}
represents the small $V$ expansion of the potential energy of a
particle in a 1D lattice in the presence of harmonic confinement when
displaced by $\delta$ from the minimum of the potential, while the
right hand side is its kinetic energy.  We can thus deduce an
approximate expression for $\delta$
\begin{equation}
  \label{eq:delta} \delta=\left[\frac{1-\cos k}{V m_{\rm
eff}}-\frac{1}{16}-\frac{\sqrt{V/J}}{256}\right]^{1/2}.
\end{equation}

Combining Eqs. \eqref{eq:COM2} and \eqref{eq:delta}, we obtain the
relation between the centre of mass oscillation frequency
$\omega_{COM}$ and the theoretical value of the mass of the different
particles
\begin{equation}
  \label{eq:1M} \omega_{COM}= V\left[\left(\sqrt{\frac{2}{m_{\rm eff}
V}}-\frac{1}{4}\right)-\frac{\sqrt{V/J}}{4} \left(\frac{1-\cos
k}{m_{\rm eff} V}-\frac{1}{16}-\frac{\sqrt{V/J}}{256}\right)\right].
\end{equation} As a benchmark, we have compared the frequency given by
Eq. \eqref{eq:1M} with the value obtained from the numerical
simulations for a free particle ($m_{\rm eff}=0.5$), and we have found
good agreement between analytical and numerical values.

\section{Dynamics of an impurity in a fermionic bath}
\label{sec:pairs-dynamics}

In the following, we characterize the different physical states which
the impurity may form inside the bath. The key observable in our
analysis is the oscillation of the time-dependent \emph{doublon
density}, defined as $\langle n_{i\,\uparrow} n_{i\,\downarrow}
\rangle (t)$, after a kick with quasi-momentum $k=0.1\pi$ has been
imparted to the impurity at $t=0^{+}$. We find that this is a more
useful quantity than the impurity density $\langle n_{i\,\downarrow}
\rangle (t)$, as it provides more direct information corresponding to
the bath dynamics and pairing with the impurity.

Examples of the oscillatory motion of the doublon density are shown in
Figs.~\ref{fig:nudN} and \ref{fig:nudN1} in the regime of strong
($|U|/J=10$) and weak ($|U|/J=1$) attraction respectively. In the
strongly attractive case, we see the oscillation frequency shifting as
a function of the bath filling, increasing only slightly while the
bath density is below half filling, but jumping to much larger values
above, from where it decreases again as the density approaches integer
filling.  Conversely, in the case of weak attraction the oscillation
frequency only decreases slightly for low and high densities, and is
roughly the same for all intermediate densities.

\begin{figure}%[!ht]
  \includegraphics[angle=-90,width=0.8\textwidth]{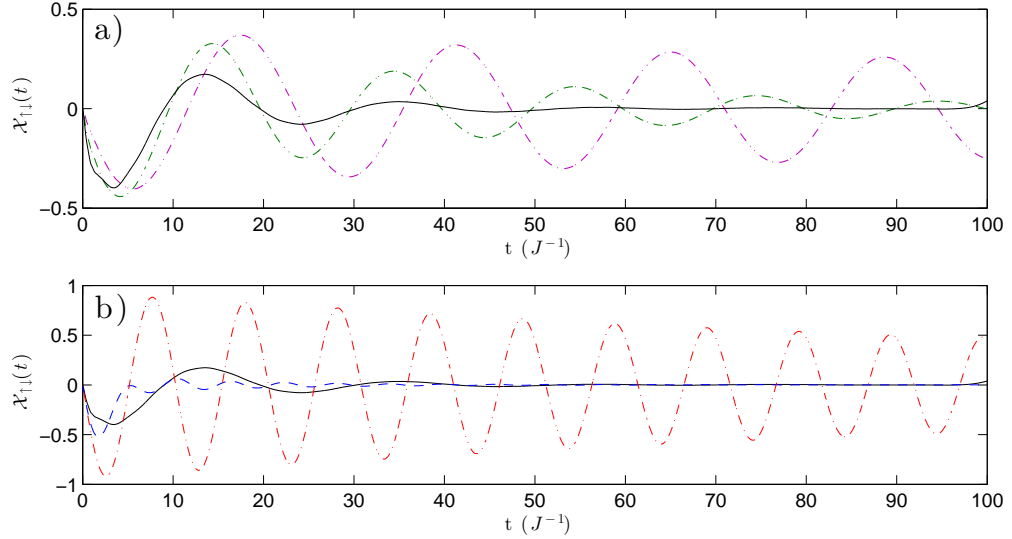}
  \caption{The doublon centre of mass $\mathcal{X}_{\uparrow\,
\downarrow}(t)$ for $|U|/J=10$ and (a) $N_\uparrow=$ 20 (purple
curve), 60 (green dash-dotted curve), 100 (black continuous curve);
and (b) $N_\uparrow=$ 100 (black continuous curve), 140 (blue dashed
curve), 160 (red curve). In panel (a), it is possible to see how the
centre-of-mass oscillations associated with the doublon dynamics are
increasingly damped when the bath population increases towards half
filling. Above half filling, the oscillations associated to the
polaron internal dynamics (described in
Sec. \ref{sec:high-frequency-peak}) start to appear, reaching the
free-particle oscillation frequency for $N_\uparrow \to L$. The
deviation of $\mathcal{X}_{\uparrow \downarrow}$ for $t \to 100
J^{-1}$ is a finite-size effect. It corresponds to the time for which
excitations in the bath are reflected back from the system boundaries
to the centre of the trap, affecting thus the doublon dynamics.  All
simulation parameters are the same as Fig. \ref{fig:nudN}.}
 \label{fig:cm_t}
\end{figure}

\begin{figure}%[!ht]
  \includegraphics[angle=-90,width=0.8\textwidth]{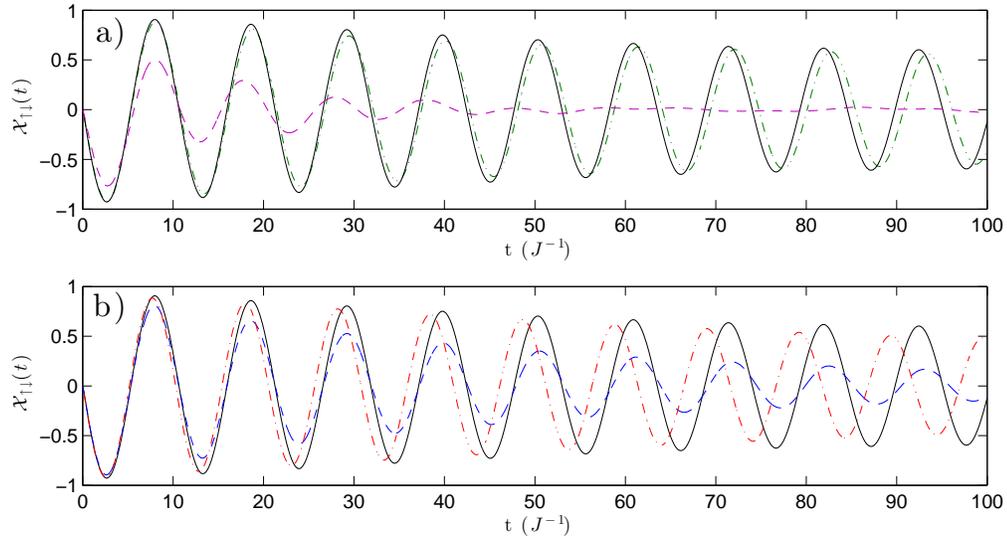}
  \caption{The centre of mass $\mathcal{X}_{\uparrow\, \downarrow}(t)$
for $|U|/J=1$ and (a) $N_\uparrow=$ 20 (purple dashed curve), 60
(green dash-dotted curve), 100 (black continuous curve), and (b)
$N_\uparrow=$ 100 (black continuous curve), 140 (blue dashed curve),
160 (red curve).  For $|U|/J=1$, the oscillations are dominated by the
single-particle oscillation frequency $\omega/J \simeq 0.62 $, and
increasingly damped for decreasing bath population.  All simulation
parameters are the same as in Fig. \ref{fig:nudN1}.}
 \label{fig:cm_t1}
\end{figure}

The observable that encapsulates all these behaviours is the doublon
centre of mass ($\mathcal{X}_{\uparrow\, \downarrow}$), defined as
$$\mathcal{X}_{\uparrow\, \downarrow}(t)=\frac{\sum_i \left(i-\frac{L-1}{2}\right) \langle n_{i\,\uparrow}n_{i\,\downarrow}\rangle(t)}{\sum_i \langle
n_{i\,\uparrow}n_{i\,\downarrow}\rangle(t)},$$ which is extracted from
the full density and is shown in Figs.~\ref{fig:cm_t} and
\ref{fig:cm_t1}.

\begin{figure}%[!ht]
  \includegraphics[angle=-90,width=0.9\textwidth]{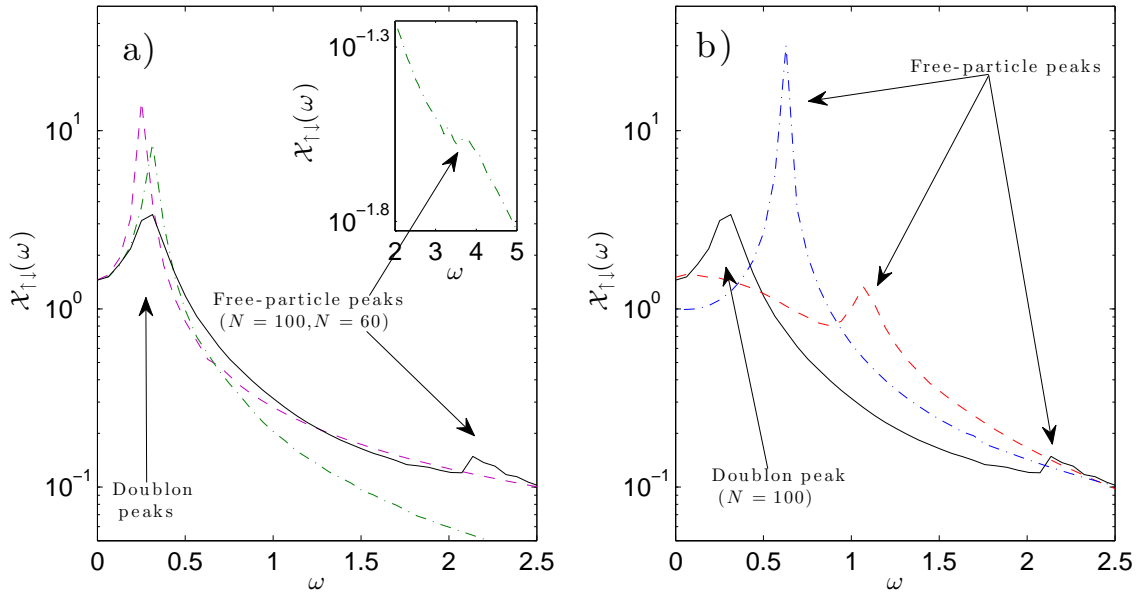}
  \caption{The doublon centre of mass $\mathcal{X}_{\uparrow\,
\downarrow}(\omega)$, $|U|/J=10$. The Fourier analysis allows a
quantitative estimation of the oscillation frequencies for different
values of $N_\uparrow$, complementing the description given in
Fig. \ref{fig:cm_t}.  (a) $N_\uparrow=$ 20, 60, 100 (purple dashed,
green dash-dotted and black continuous curves). For $N_\uparrow=$ 20
only the doublon peak for $\omega/J \simeq 0.5$ is present, while for
$N_\uparrow=$ 100 the polaron internal dynamics peak appears, for
$\omega/J \simeq 2.1$ respectively, and for $N_\uparrow=$ 60 the
precursor of the polaron peak starts to appear around $\omega/J \simeq
3.8$. The inset shows additional data for higher values of $\omega$
. (b) $N_\uparrow=$ 100, 140, 180 (black continuous, blue dashed, and
red dash-dotted curve). The polaron internal dynamics peak shifts to
the left for increasing interaction ($\omega/J \simeq$ 2.1, 1.1, 0.8
for $N_\uparrow=$ 100, 140, 180), while the doublon peak is
increasingly damped. All simulation parameters are the same as in
Fig. \ref{fig:nudN}. }
 \label{fig:cm_w}
\end{figure}

\begin{figure}%[!ht]
  \includegraphics[angle=-90,width=0.9\textwidth]{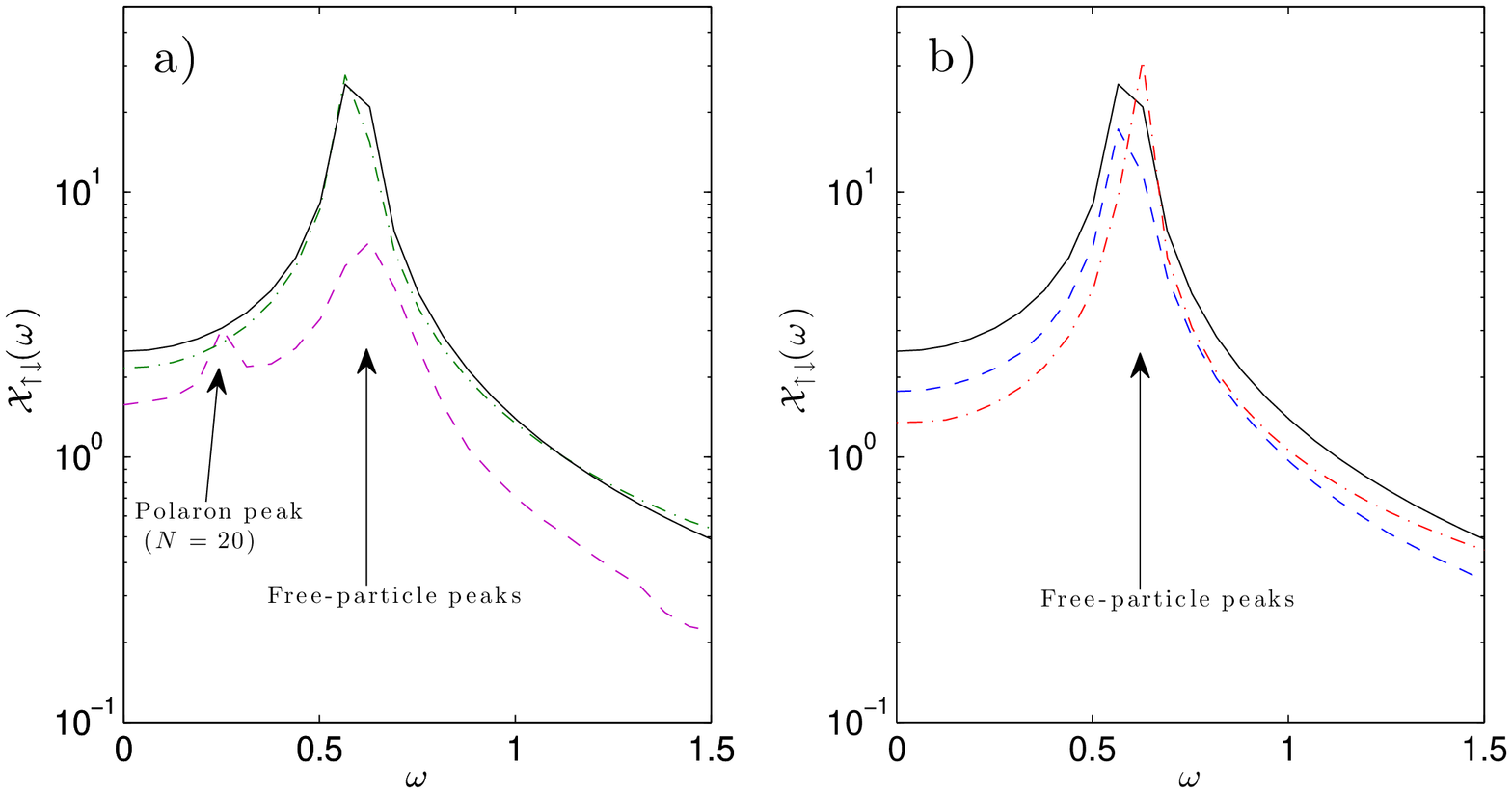}
  \caption{$\mathcal{X}_{\uparrow\, \downarrow}(\omega)$,
$|U|/J=1$. (a) $N_\uparrow=$ 20, 60, 100 (purple dashed, green
dash-dotted and black continuous curves) and (b) $N_\uparrow=$ 100,
140, 180 (black continuous, blue dashed, and red dash-dotted
curve). For $N_\uparrow=$20 both the polaron and the free particle
peaks are present ($\omega /J \simeq$ 0.4, 0.8 respectively, while for
larger values of $N_\uparrow$, only the free particle peak is present.
All simulation parameters are the same as Fig. \ref{fig:nudN1}.}
 \label{fig:cm_w1}
\end{figure}

As is explained in detail in the following sections for the different
interaction and bath density regimes, we gain insight into the physics
of the system by analysing the Fourier transform of
$\mathcal{X}_{\uparrow\, \downarrow}(t)$, $\mathcal{X}_{\uparrow\,
\downarrow}(\omega)$, which is shown in Figs. \ref{fig:cm_w} and
\ref{fig:cm_w1} for strong and weak attraction. For strong attraction,
$\mathcal{X}_{\uparrow\, \downarrow}(\omega)$ shows that the
oscillation of a tightly bound on-site pair dominates the dynamics at
low density, while a polaron-like state is present but weak (low- and
high-frequency peaks in Fig. \ref{fig:cm_w}-a respectively). The
relative weights of the polaronic and bound-pair peaks reverse as the
density of bath atoms increases above $0.5$, with the polaron
component becoming predominant and the bound-pair peak almost
vanishing (high- and low-frequency peaks in Fig. \ref{fig:cm_w}-b
respectively; notice e.g. how for $N_{\uparrow}=140$ the bound-pair
peak has become essentially just a broad shoulder).

In order to understand the physics of the bound pair for strong
interactions between impurity and bath atoms, we make use of the
so-called string hypothesis from the Bethe-ansatz solution of the
Hubbard model (c.f. Appendix \ref{app:bethe}). Using this, we see that
a tightly-bound on-site pair dominates at low bath fillings, and can
be understood in a two-body picture, where we can compute formulas for
the pair oscillation frequency, as shown in Fig. \ref{fig:mass}. On
the other hand, the polaron-like state dominating at higher densities
can be explained by the scattering of bath particles from the Fermi
surface to the edge of the Brillouin zone meditated by the oscillating impurity
. We find that the frequency of the resultant peak in
$\mathcal{X}_{\uparrow\, \downarrow}(\omega)$ is insensitive to the
value of both interaction strength $U$ and initial momentum kick
$k$. We also examine the way the impurity modifies the density of the
bath around its position and explain why the peak position stays
independent of $U$ despite this local distortion. Subsection
\ref{sec:high-frequency-peak} describes this in detail, including the
change in the polaron oscillation frequency with density.

By contrast, in the case of weak interaction, a polaron component to
the oscillation is significant only for densities at or below $0.2$
(low-frequency peak in Fig. \ref{fig:cm_w1}-a), while the dominant
component of $\mathcal{X}_{\uparrow\, \downarrow}(\omega)$ stems from
the motion of a free particle (low- and high-frequency peaks in
Figs. \ref{fig:cm_w1}-a and b, respectively). In this regime the
polaronic component corresponds to the resonant scattering between a
bound pair and an impurity particle propagating through the background
particle bath.  This will be described in section \ref{sec:weak}.

\subsection{Strong interactions: the low-frequency peak}
\label{sec:low-frequency-peak}

In a 1D Hubbard model with strong interactions, Bethe-ansatz
techniques provide two sets of solutions, which together cover all
possible eigenstates of the system - provided we assume the so-called
string hypothesis to be correct \cite{Essler:2005uw}.  In each of
these two sets of solutions, A and B, the total energy $E$ and
quasi-momentum $P$ of a many-body state can be expressed in terms of
Bethe-ansatz quantum numbers $\{k_i\}$ and $\Lambda$.

For the solutions of type A
  \begin{equation}
    \label{eq:8aT} P = \sum_{j=1}^{N} k_j \qquad E = -2 J \sum_{j=1}^N
\cos\left(k_j\right)+ const.
  \end{equation} and for the solutions of type B,
\begin{equation}
  \label{eq:10aT} P = \sum_{j=1}^{N-1} k_j + 2 q \qquad E = -2 J
\sum_j^{N-1} \cos\left(k_j \right) - 4 J \cosh \xi \cos \left( q
\right) + const.
\end{equation} where $\xi$ is defined by
\begin{equation}
  \label{eq:11a2} \cosh \xi = \sqrt{1+\frac{U^2}{16 J^2
\cos^2\left(q\right)}}
\end{equation} and $q$ is the real part of the quantum numbers
associated with the $k-\Lambda$ string (see Appendix \ref{app:bethe}).
In the strong coupling limit, it is possible to show that
\begin{equation}
  \label{eq:12a2} -4 J \cosh \xi \cos q \to U -4 J^2 / \left| U
\right| -4 J^2 / \left| U \right| \cos \left( \kappa \right)
+O\left(\frac{1}{U^2}\right),
\end{equation} where $\kappa = 2 q$ is the total quasi-momentum of the
pair.

The eigenstates of type A correspond to the effectively free motion of
both bath particles and the impurity, while solutions of type B
correspond to a bound state of the impurity with one bath atom (with
the remaining $N_{\uparrow}-1$ bath atoms again moving freely) (see
Appendix \ref{app:bethe}).  For an attractive interaction, the B-type
solutions are always energetically favorable, and for $|U|/J>2$ there
is no overlap between the bands associated with type-A and type-B
solutions.

Thus focussing on the B-type manifold of eigenstates, we can describe
the observed behaviour of the low-frequency peak in
$\mathcal{X}_{\uparrow \downarrow}(\omega)$ at large $|U|$
(c.f. Fig. \ref{fig:cm_w}) by deriving the explicit expression for the
many-body energy $E$ in Eq. (\ref{eq:10aT}) as a function of the
doublon quasi-momentum $\kappa$ (see Appendix \ref{app:bethe}) to be
\begin{equation}
  \label{eq:Edoubl} E_{\kappa}=-\sqrt{U^2+16 J^2
\cos^2\left(\kappa/2\right)} ,
\end{equation} where the dependency of the energy on the other
$N_{\uparrow}-1$ quasi-momentum quantum numbers $\{k_i\}$ has not been
taken into account, being irrelevant in the dynamics here considered.
From this, the effective mass of the tightly bound pair can be
extracted to be
\begin{equation}
  \label{eq:2} m_{\rm doublon}=\left[ \left(\frac{\partial^2
E}{\partial \kappa^2}\right)\right]^{-1}_{\kappa \to
0}=\frac{1}{J}\sqrt{1+\frac{U^2}{16 J^2}} \xrightarrow{|U|/J \gg 1}
\frac{|U|}{4 J^2},
\end{equation} which, in the limit $|U|/J\ll 1$, coincides with the
expression that would be obtained from second-order perturbation
theory.

\begin{figure}%[!ht]
  \includegraphics[width=0.6\textwidth]{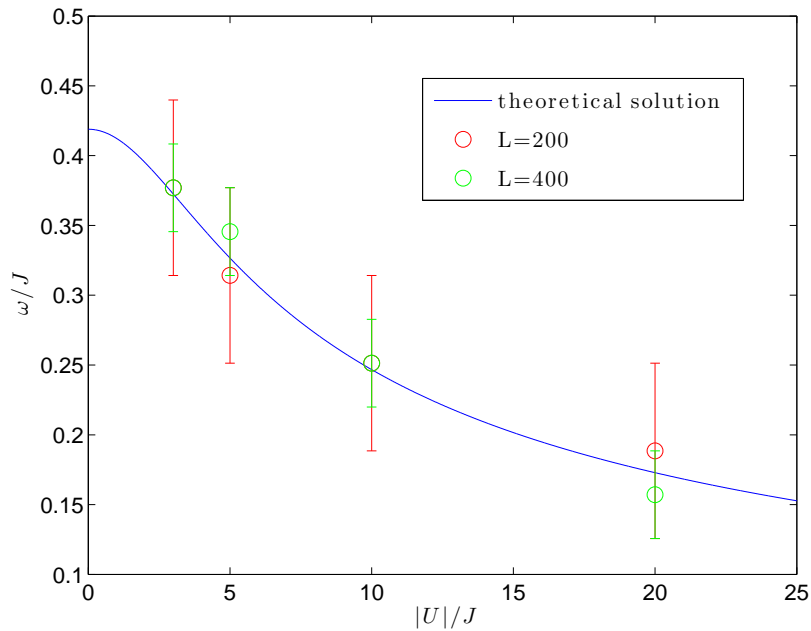}
  \caption{Comparison between the numerical value of the low-frequency
peak oscillation frequency for the large interaction case (as
extracted from $\mathcal{X}_{\uparrow \downarrow}(\omega)$), and the
theoretical value for $|U|/J>1$, given by
Eqs. \eqref{eq:Edoubl}~\eqref{eq:2}). This comparison allows us to
establish that the low-frequency oscillations of
$\mathcal{X}_{\uparrow\, \downarrow}$ can be described in terms of
oscillations of a particle (composed by one spin-up and one spin-down
particle) in a parabolic confining potential.  All simulation
parameters are the same as in Fig. \ref{fig:nudN}, except the
interaction strength and the lattice size. Here $|U|/J=$3, 5, 10, 20
and $L=$ 200 or 400.  The error bars account for the finite resolution
of the Fourier transform. Larger lattices allow for longer simulation
times before the reflection from the edges starts playing any role in
the doublon dynamics, thus allowing higher resolutions in the
frequency domain. It is worth noting that the full Bethe-ansatz
formula works even for moderately weak interaction. The characteristic
size of the composite particle is given by $\xi^{-1}$ (see
Fig. \ref{fig:psize}), corresponding to a value of $\simeq 2.3$ for
$|U|/J=3$.}
 \label{fig:mass}
\end{figure}

It is interesting to note that our Bethe ansatz solution delivers the
same result for the quasi-momentum dependence of the doublon
contribution to the energy as the simple solution to the problem of
two distinguishable bound particles on a lattice does
\cite{Winkler:2006wu,Valiente:2008kb}. In Fig.~\ref{fig:mass} we show
the comparison between the numerical value of the doublon peak
oscillation frequency (low-frequency peak in the strong-interaction
regime), and the theoretical value obtained using our value for the
doublon mass \eqref{eq:2} in conjunction with Eq. \eqref{eq:1M}. The
excellent agreement shows that the doublon dynamics is well-captured
by our Bethe-ansatz based model.

\subsection{Strong interactions: the high-frequency peak}
\label{sec:high-frequency-peak}

In the case of strong attraction (large $|U|$) treated in this
section, the oscillations of the doublon density also show another,
high-frequency component in $\mathcal{X}_{\uparrow \downarrow}(t)$,
which is weak for low bath densities, but which becomes significant
above half-filling (see Fig.  \ref{fig:cm_w}). Here, we argue that
this feature can be understood in terms of an effective scattering of
bath particles from the Fermi surface off the oscillating impurity
towards the edge of the Brillouin zone.  We will show that the
position of this high-frequency peak in $\mathcal{X}_{\uparrow\,
\downarrow}(\omega)$ depends primarily on the filling fraction in the
bath, and the associated value of $k_F$, and is essentially
independent of the kick strength and insensitive to $U$.

\subsubsection{Description of the dynamics in terms of particle-hole
excitations of the bath}

\begin{figure}%[!ht]
  \includegraphics[angle=0,width=0.35\textwidth]{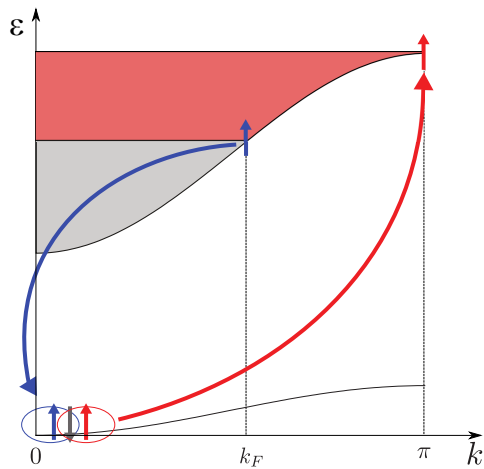}
  \caption{Two-band process leading to the scattering of a bath
    particle from $k_F$ to $\pi$. The specific pair-mediated process
    depicted here ($k_P=\pi$) represents the process associated with
    the maximum energy transfer from the pair to the bath. The
    promotion of an $\uparrow$ particle from the pair to the bath is
    allowed for $k_F<k <\pi $, while for $0<k<k_F$ the process is forbidden 
    due to Pauli blocking.}
   \label{fig:2band}
\end{figure}

The high-frequency peak in the strongly interacting limit can be
understood in terms of scattering between two bath particles mediated
by the presence of the impurity, together with the dynamics of the
impurity itself.  Within the framework of a two-band model (see
Fig. \ref{fig:2band}), the effective scattering between bath particles
is explained in terms of an exchange process, involving the transfer
of the up particle from the tightly bound pair to the
$\uparrow$-particle band above the Fermi level, and the concomitant
transfer of a particle from the Fermi surface to the tightly bound
pair (see Fig. \ref{fig:2band}).

 The total energy necessary for this process, which the
initial kick must supply now, involving both the scattering process as
well as the impurity dynamics, is given by
\begin{equation}
  \label{eq:hi_pk} \Delta E = \omega_{hi} + 2 J \left[ (1-\cos k_p) -
(1-\cos k_{F\,\uparrow})\right],
\end{equation} where $\omega_{hi}$ is the oscillation energy
associated with the dynamics of the impurity in a completely full
bath-particle band. The frequency $\omega_{hi}$ thus corresponds 
to the oscillation frequency of a free particle in the lattice, in 
presence of the parabolic confining potential $V$, and its value can be 
calculated from Eq. \eqref{eq:1M}. 
 
The term $2J(1-\cos k_p)$ corresponds to the transfer of an
''$\uparrow$'' particle from the tightly bound pair to a free state of
the bath above the Fermi level, as sketched in
Fig. \ref{fig:2band}. The term $(1-\cos k_{F\,\uparrow})$ then is
related to the transfer of an $\uparrow$-particle from the Fermi
surface of the bath to the bound state of the $\uparrow\downarrow$-
pair.

\begin{figure}%[!ht]
  \includegraphics[angle=0,width=0.5\textwidth]{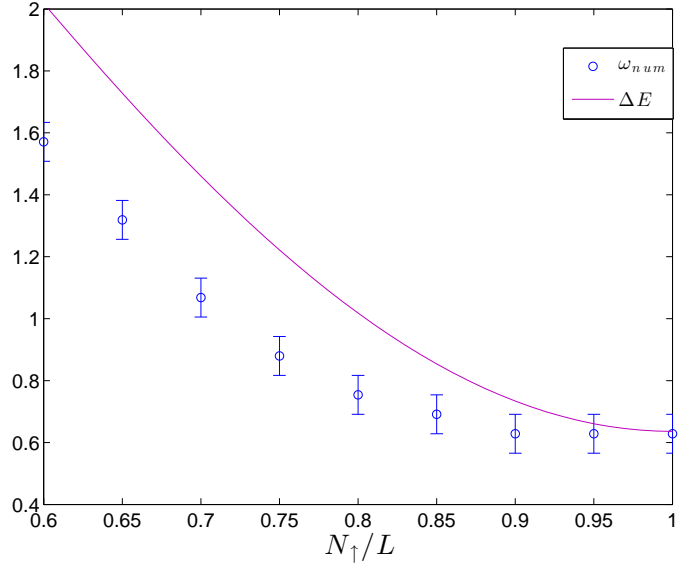}
  \caption{ Plot of the high-frequency peak as a function of the bath
filling, compared to the value of $\Delta E$ obtained from
Eq. \eqref{eq:hi_pk}, for $k_p=\pi$. The position of the peaks in the
numerical results does not show any dependence on $k$ and $U$: the
value of $\omega_{num}$ coincides for $k=0.05 \pi,\, 0.1 \pi,\, 0.25
\pi$ and $|U|=10$, and $k=0.1\pi$ and $|U|=5$.}
 \label{fig:cross}
\end{figure}

In Fig. \ref{fig:cross}, we show how the position of the
high-frequency peak depends on the bath density - provided by numerics
- and that it is in reasonable agreement with the expression given by
Eq. \eqref{eq:hi_pk} when $k_p=\pi$, corresponding to the largest
possible energy associated with the transfer of an $\uparrow$-particle
from the pair to the bath (see Fig. \ref{fig:2band}).  We further find
the peak position to be insensitive to changes in $U$ and kick
strength $k$.

The offset between numerical results and eq. \eqref{eq:hi_pk} is then
related to the nonuniform spatial distribution of $k_F$ for the bath
particles. This nonuniformity, in turn, is due to the perturbing
effect of the impurity, which can be approximated as Friedel
oscillations in the bath as we will show below.  As the spatial extent
of Friedel oscillations does not depend on $|U|$, the spatial
distribution of $k_F$ and consequently $\Delta E$ are independent of
the strength of the interaction. These findings go some way towards
explaining why even the local density disturbance of the bath by the
impurity is congruent with the observed $U$-independence of the
high-frequency peak.

\subsubsection{Bath-particle distribution in presence of an impurity}

\begin{figure}%[!ht]
  \includegraphics[angle=-90,width=0.8\textwidth]{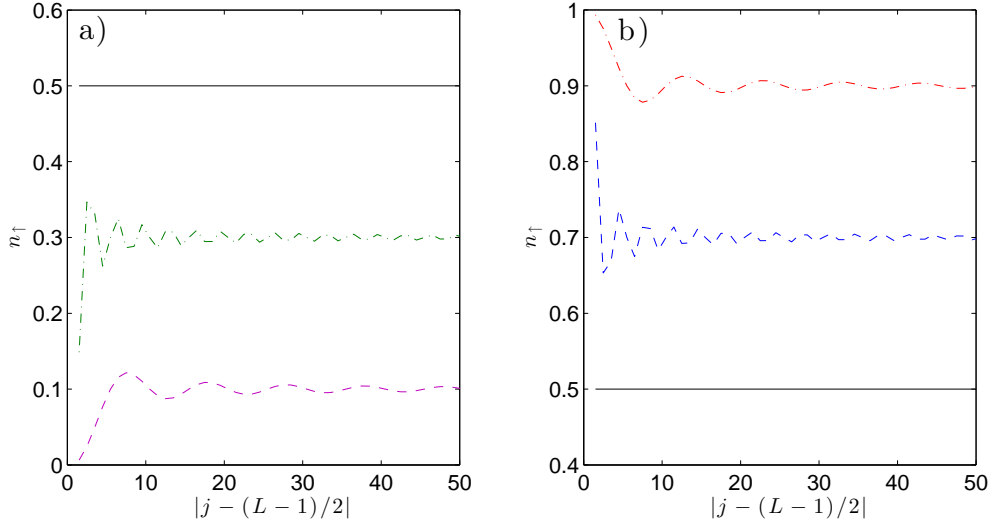}
  \caption{Friedel oscillations (see Eq. \eqref{eq:5}) induced by the
presence of a hard-wall boundary condition at $j=(L-1)/2$ for (a) 20
(purple dashed curve), 60 (green dash-dotted curve), 100 (black
continuous curve), and (b) 100 (black continuous curve), 140 (blue
dashed curve), 180 (red dash-dotted curve) bath particles respectively
(see Eq. \eqref{eq:5}).}
 \label{fig:Fried}
\end{figure}

With attractive interactions and an impurity that is localized by the
tight parabolic confinement potential, the density of bath particles
is modified locally in the center of the system. Let us denote the
size of this modified region of the bath by $\zeta$. We make a first
estimate for the value of $\xi$ by considering the limiting case $U
\to -\infty$, with the impurity being located on a single site,
$i$. In this case, due to the limit of infinite attraction, the site
$i$ will act as a hard-wall boundary condition
\cite{Matveev:1993jy,Fabrizio:1995ta}, and the density profile of the
bath particles around the impurity will undergo Friedel oscillations
\cite{Friedel:1958vd} (see Fig. \ref{fig:Fried}), as was recently
pointed out in \cite{Lamacraft:2009bp} for repulsive interaction and
in absence of the lattice, according to the following formula
\begin{equation}
  \label{eq:5}
\rho_\uparrow(j)=\frac{N_\uparrow}{L}-\frac{1}{2\pi}\frac{\sin\left[2
k_{F\,\uparrow}\left(j-i\right)\right]}{|j-i|} .
\end{equation} We note that in a 1D system the period of these
oscillations is always independent of the strength of the interactions
between the bath particles and the bath impurity interactions, and is
\textit{always} $(2 k_{F\,\uparrow})^{-1}$ for a fermionic bath, and
$(2 \pi \rho_0)^{-1}$ for a bosonic one - all that changes with the
interaction is the amplitude of the Friedel oscillations.

\begin{figure}%[!ht]
  \includegraphics[angle=-90,width=0.8\textwidth]{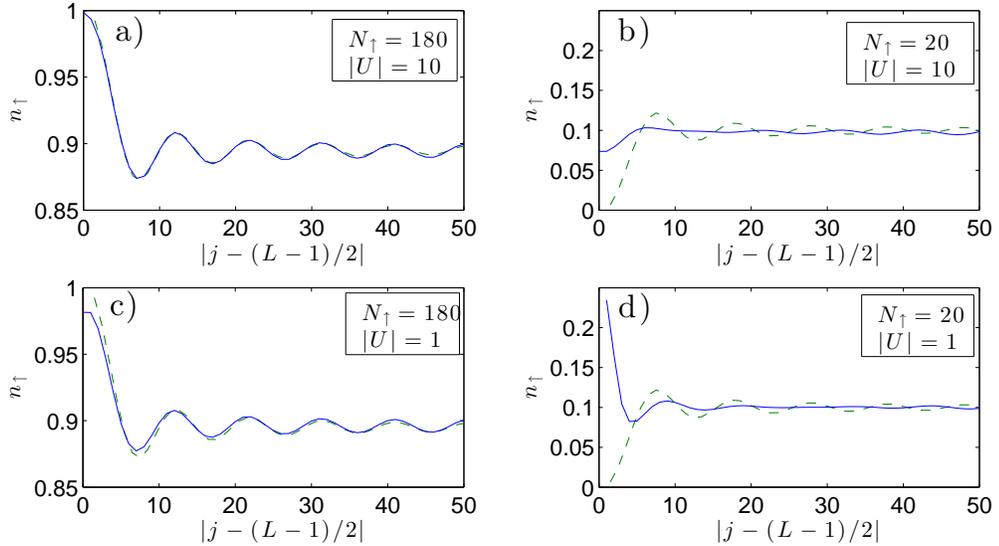}
  \caption{ Comparison between the numerical estimate of the
ground-state spatial oscillations induced in the bath by (a,c,d) a
single down impurity or (b) a doublon (blue continuous curves), with
the Friedel oscillations induced by hard-wall boundary conditions
(green dash-dotted curves), with (a) $N_\uparrow=180$, $|U|/J=10$, (b)
$N_\uparrow=20$, $|U|/J=10$, (c) $N_\uparrow=180$, $|U|/J=1$, (d)
$N_\uparrow=20$, $|U|/J=1$.  In the high-density limit, the agreement
between the numerical calculations and the approximate model is good,
while, in the low-density limit, the numerical results suggest that
the role of the impurity cannot be described in terms of an
impurity-induced Hartree potential. This hypothesis is confirmed by
the exact diagonalisation of an Hamiltonian describing a
spin-polarized gas in presence of a localized potential mimicking the
potential induced by the impurity. In this case the results are
compatible with the results implied by Eq. \eqref{eq:5}.  All
simulation parameters are the same as in Fig. \ref{fig:nudN} and
Fig. \ref{fig:nudN1}.}
 \label{fig:Fried_U10}
\end{figure} 

In Fig. \ref{fig:Fried_U10}, we show how approximating
either the on-site bound pair (in case of strong attraction) or the
single particle (for weak attraction) as a boundary condition located
at the minimum of the parabolic potential works well to describe the
density modulation induced in the bath when $N_\uparrow$ is large,
while it fails in the case of a more dilute system. As we have been
focussing on just this regime of intermediate and high filling in this
part, this approximation should be very reasonable.

\subsection{Weak interactions}
\label{sec:weak} 
\begin{figure}%[!ht]
  \includegraphics[angle=-90,width=0.5\textwidth]{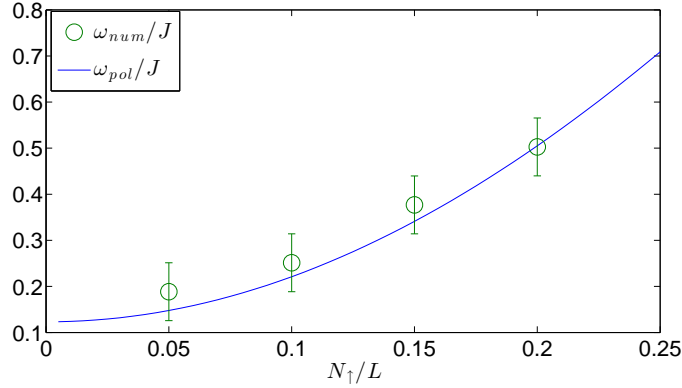}
  \caption{Comparison between the numerical value of the low-frequency
peak at $|U|/J=1$ and $\omega_{pol}$ corresponding to the resonant
frequency of the bound-doublon to scattering-states transition. This
peak can be seen in the $N_\uparrow=20$ plot (small peak around
$\omega \simeq 0.2$ in Fig. \ref{fig:cm_w1}).  Above $N_\uparrow/L
\simeq 0.2$ it merges with the free-particle peak, and eventually
disappears.  In this picture, the spin-up bath particles are
resonantly bound to the spin-down impurity, analogous to the case of
static polarons. All simulations parameters are like in
Fig.~\ref{fig:nudN1}. The errorbars account for the finite resolution
of the Fourier transform.}
 \label{fig:pol_pk}
\end{figure}

As anticipated, for interactions $|U|\leq 1$, the
peak in $\mathcal{X}_{\uparrow\,\downarrow}(\omega)$ associated with
the oscillation of a tightly bound on-site pair is not
present. Nevertheless, we observe two distinct modes in the
doublon-density center-of-mass oscillations
$\mathcal{X}_{\uparrow\,\downarrow}(t)$ in the weakly interacting
regime as well.  One of them appears due to the free oscillatory
motion of a non-interacting particle as derived in
Sec. \ref{sec:free-particle}.  Another peak, at lower frequencies,
appears as well (see Figs. \ref{fig:cm_w1} and \ref{fig:pol_pk}).

The underlying physics of this low-frequency peak derives from a
resonant transition between a bound pair and scattering states,
specifically the spin-down impurity at zero quasi-momentum and a
spin-up bath particle at the Fermi quasi-momentum $k_{F
\uparrow}$. The frequency of this peak can thus be obtained
considering the difference between the energy of the (weakly-bound)
pair and the energy of the scattering state
\begin{align}
  \label{eq:3} \omega_{pol}=
E_{\uparrow}+E_{\downarrow}-E_{\uparrow\,\downarrow}= -2 J \left[1+
\cos(k_{F\,\uparrow})\right]+\sqrt{U^2+16 J^2} .
\end{align} Figure \ref{fig:pol_pk} shows the excellent agreement
between this model and the numerical results from the TEBD
calculations, within the error bars set by the finite resolution of
the Fourier transform.

Further, the existence of the two peaks in
$\mathcal{X}_{\uparrow\,\downarrow}(\omega)$ can be related to the
known structure of the polaron ground state: in the the polaron ansatz
\cite{Chevy:2006hv} of the type \begin{equation}
 \label{eq:pol} | \Psi \rangle = \sqrt{Z} c^\dagger_{0\downarrow} | FS
\rangle_\uparrow | 0 \rangle_\downarrow +
\sum_{k>k_F^\uparrow,q<k_F^\uparrow} \phi_{k,q} c^\dagger_{k\uparrow}
c_{q\uparrow} c^\dagger_{q-k \downarrow} | FS \rangle_\uparrow | 0
\rangle_\downarrow,
\end{equation} the first term describes the impurity at rest in the
presence of the unperturbed Fermi sea of the bath (with the
quasiparticle weight $Z$), and the second term a coherent
superposition of states, in which the motion of the impurity is
correlated with a single particle-hole excitation of the Fermi sea
(like the weakly bound state entering Eq. (\ref{eq:3})).

We can hypothesize that the high-frequency ``free particle'' peak in
$\mathcal{X}_{\uparrow\,\downarrow}(\omega)$ corresponds to a
second-order process involving the virtual breaking of a pair, while
the low-frequency peak is caused by breaking up the correlated state
between impurity and a particle-hole excitation of the Fermi sea.

% When such a state experiences an initial quasi-momentum kick at
$t=0^+$, both parts of Eq. (\ref{eq:pol}) % will contribute to the
dynamics. Based on our results, we can hypothesize that the first term
of Eq. (\ref{eq:pol}) leads to % the high-frequency "free particle"
peak in $\mathcal{X}_{\uparrow\,\downarrow}(\omega)$, while the second
term causes % the low-frequency peak, which is caused by breaking up
the correlated state between impurity and a particle-hole % excitation
of the Fermi-sea, as we have shown through the agreement of
Eq. (\ref{eq:3}) with our simulation data % (c.f. Fig.
\ref{fig:pol_pk}). The relative height of the peaks could perhaps
provide a useful tool for determining the % relative weights of the
two terms in Eq. (\ref{eq:pol}).

\subsection{Damping}
\label{sec:peak-damping}

Along the lines of the above discussion about the particle-hole
excitation process, it is possible to intuitively understand the
oscillation damping. When approaching half-filling, both from the low-
and the high-density limit, the oscillation damping is increased. This
increase is associated with the increase of the particle-hole creation
mechanism through the virtual breaking of a pair for increasing
filling, and the concomitant energy transfer increase for decreasing
filling. This mechanism of dissipation is confirmed by the observation
in the numerical data of density perturbations in the bath particles,
propagating at $2J$, consistent with the picture of the transfer of
bath particles to the top of the band $k=\pi$.

\section{The kicked impurity in a bosonic reservoir}
\label{bosons}

Comparing the results obtained for a fermionic reservoir to those from
a bosonic one enables us to state which features of the observed
dynamics are universal, and which are particular to the fermionic
bath. Towards this, we have performed TEBD simulations for a
two-species Bose-Hubbard Hamiltonian
\begin{equation}
   \label{eq:BHubb} H= -J\sum_{i \, \sigma} b_{i \sigma}^\dagger
b_{i+1 \sigma} + h.c. + U \sum_i n_{i\, \uparrow}n_{i\, \downarrow} +
\frac{W}{2} \sum_i n_{i\, \uparrow}(n_{i\, \uparrow}-1) + V \sum_i
n_{i \,\downarrow} \left(i-\frac{L-1}{2}\right)^2,
   \end{equation} where trap parameter $V$, tunneling $J$ and the
values of $U$ and $N_{\uparrow}$ are the same as described in section
\ref{fermions}, as is the impurity preparation and the initial kick to
the impurity (as in the fermionic case we will set $J=1$, $a=1$). The
two key differences are that now $b_{i \, \sigma}$, $b_{i \,
\sigma}^{\dagger}$ are operators for softcore bosons, which interact
repulsively on-site with energy $W$, if $\sigma=\uparrow$. Here, we
have considered $W/J=4,10, 20$.

\subsection{The weak interaction limit}

\begin{figure}%[!ht]
  \includegraphics[width=0.8\textwidth]{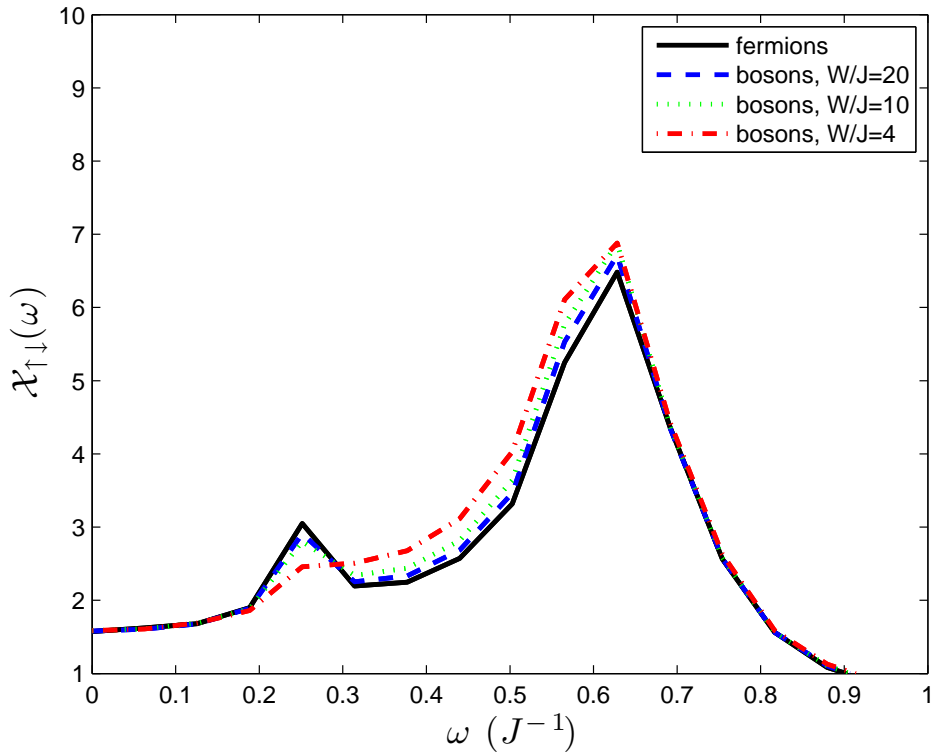}
  \caption{Comparison of $\mathcal{X}_{\uparrow\,\downarrow}(\omega)$
between fermions and bosons in the weak-attraction case, $|U|/J=1$,
$N_{\uparrow}=20$. The position of the low-frequency peak for the
fermionic bath, which stems from the break-up of correlated states
between the impurity and particle-hole excitations in the bath
(c.f. \ref{sec:weak}), is very similar for the bosonic bath, at all
values of $W$. See text for details.}
 \label{fig:comp_fermi_bose}
\end{figure}

Repeating the TEBD simulations for weak interactions between the
impurity and a bosonic bath, we find that the two-peak structure
discussed in \ref{sec:weak} persists for all values of $W$ we have
simulated, as shown in Fig. \ref{fig:comp_fermi_bose}. Moreover, the
position of the polaron-dissociation peak (c.f. \ref{sec:weak}) is
still described remarkably well by the theory for the fermionic bath,
even for the lowest value of $W$, $W/J=4$.  These findings can be
understood analytically by observing that, at low densities and
reasonably large values of $W$, one-dimensional lattice bosons map to
spinless fermions with weak nearest neighbor attractions.  This
mapping is achieved by describing the sector of low-energy,
long-wavelength excitations of the $\uparrow$ component of the
Hamiltonian (\ref{eq:BHubb}) as a Tomonaga-Luttinger liquid (TLL),
whose properties are characterized by the so-called TLL parameters,
$K_b$ and $v_b$ \cite{Cazalilla:2011}.  Expanding the
$\uparrow$-sector in Hamiltonian (\ref{eq:BHubb}) and the bath density
operator $n_{x\, \uparrow}$ in terms of the canonically conjugate TLL
field operators $\phi_{\uparrow}(x)$ and $\theta_{\uparrow}(x)$, one
obtains
\begin{eqnarray}
  \label{eq:HubbTLL} H & \approx &\frac{1}{2\pi}\int dx
\left(u_bK_b(\partial_x\theta_{\uparrow}(x))^2+\frac{u_b}{K_b}(\partial_x\phi_{\uparrow}(x))^2\right)+U\sum_i
\left(\rho_0-\frac{\partial_x\phi_{\uparrow}(x_i)}{\pi}\right)\left(\sum_{m=-\infty}^{\infty}
e^{2im [\pi\rho_0x_i- \phi_{\uparrow}(x)]}\right)n_{i\, \downarrow}
\nonumber \\ && + J\sum_{i} b_{i \downarrow}^\dagger b_{i+1
\downarrow} + h.c. + V \sum_i n_{i \,\downarrow}
\left(i-\frac{L-1}{2}\right)^2,
\end{eqnarray} where expanding the impurity-bath coupling in this way
presupposes that the impurity does not distort the bath density too
much locally. In the limit of large $W/J$, the TLL parameters are
known perturbatively, $K_b\approx 1+\frac{4J}{\pi W}\sin\pi\rho_0$,
$u_b=2 J \sin\pi\rho_0\left(1-\frac{4 J}{W}\rho_0\cos\pi\rho_0\right)$

Now, a bath of spinless lattice fermions with attractive
nearest-neighbor interactions $V_{nn}$ and Fermi quasi-momentum $k_F$
coupled to an impurity with an on-site density-density interaction,
can be mapped to a TLL in 1D in a manner identical to
(\ref{eq:HubbTLL}), where $K_b$ is replaced by $K_f$ and $u_b$ by
$u_f$ \cite{Giamarchi:2003}. These parameters are also known from
perturbation theory for small $V_{nn}$:
$K_f=1+\frac{V_{nn}(1-\cos2k_F)}{2\pi \sin k_F}$, $u_f=2 J\sin
k_F\left(1-\frac{V_{nn}(1-\cos2k_F)}{2\pi J \sin k_F} \right)$.

Computing $K_b$ and $u_b$ for densities between $0$ and $0.2$ and
large $W/J$ shows that Eq. (\ref{eq:HubbTLL} ) can be read
equivalently as the model of an impurity coupled to weakly
nearest-neighbour attractive spinless fermions. For example, for
$W/J=10$ and $n_{\uparrow}=0.1$, $K_b=1.039$, $u_b/J=0.59$, values
that are best matched by $K_f$ and $u_f$ for $|V|/J=0.75$,
$k_F=0.1\pi$.  Crucially, a value of $K_f=1.039$ is still very close
to the values for free fermions, $K_f=1$. Thus, the continued
applicability of Eq. (\ref{eq:3}) - which had initially been developed
from a one-body picture of the free fermion bath - to predict the
polaron peak in the weak-coupling regime even for a (sufficiently
repulsive) bosonic bath, can be explained
(c.f. Fig. \ref{fig:comp_fermi_bose}).

\subsection{The strong interaction limit - higher-order bound states
of the impurity.}  
\begin{figure}%[!ht]
  \includegraphics[width=1.0\textwidth]{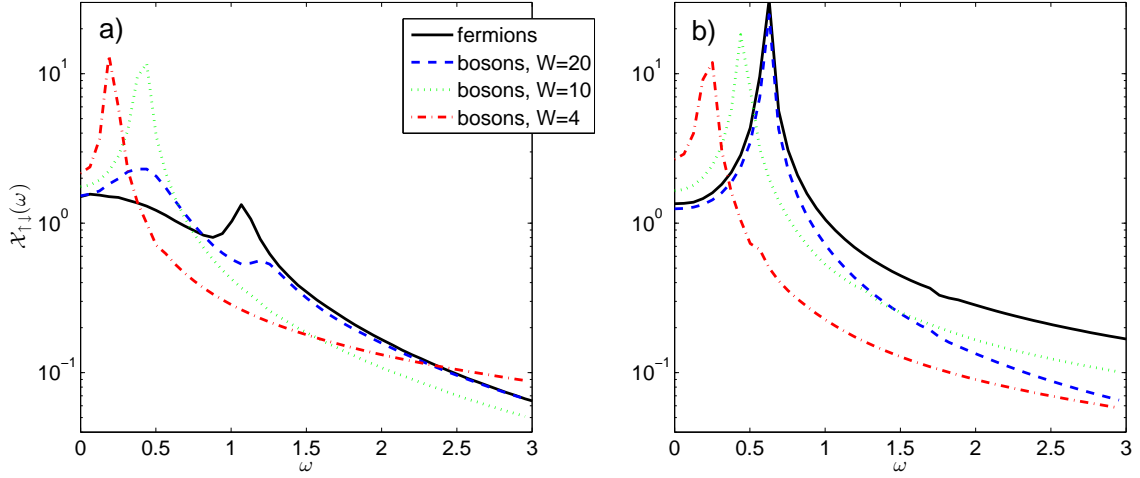}
  \caption{Comparison of $\mathcal{X}_{\uparrow\,\downarrow}(\omega)$
between fermions and bosons in the strong-attraction case,
$|U|/J=10$. a) $N_{\uparrow}=140$. For the fermion bath (black solid
line), the virtual pair-breaking peak
(c.f. \ref{sec:high-frequency-peak}) has become dominant, with the
doublon peak at lower frequency having almost completely flattened
out. For bosons at $W/J=20$ (blue dashed line), the behaviour is
qualitatively similar, but the oscillation of the doublon is still
dominant, as evidenced by the larger low-frequency peak. For $W/J=10$
(green dotted line) and $W/J=4$ (red dash-dotted line), only the
high-order bound states oscillate, there is no virtual pair-breaking
peak.  b) $N_{\uparrow}=180$. For the fermion bath, the virtual
pair-breaking peak (c.f. \ref{sec:high-frequency-peak}) is now fully
dominant, with the doublon peak at lower frequency having completely
vanished. The dynamics for the boson bath at $W/J=20$ (blue dashed
line), now almost completely matches that of the fermions, showing
that for bosons as well virtual pair-breaking peak, corresponding to
the fermionization of the bosonic gas. For $W/J=10$ and $W/J=4$, only
the high-order bound states oscillate as in a), there is again no
virtual pair-breaking peak.}
 \label{fig:comp_fermi_bose2and3}
\end{figure}

In the opposite limit of large bath-impurity
attraction, $|U|\gg J$, the kick-induced dynamics of the impurity
start to depend crucially on the ratio $|U|/W$. As an example, as for
the fermions, we focussed on the case $|U|/J=10$.  As long as $W>|U|$,
the spectrum of doublon dynamics,
$\mathcal{X}_{\uparrow\downarrow}(\omega)$, remains qualitatively
unaltered from the case of the fermionic bath: when
$n_{\uparrow}<0.5$, the dynamics of the kicked impurity are dominated
by the oscillation of the doublon-mode, whereas for
$n_{\uparrow}\geq0.5$ this doublon peak increasingly flattens out and
eventually disappears as $n_{\uparrow}$ increases above half-filling,
as shown in Fig. \ref{fig:comp_fermi_bose2and3}. At the same time,
like for fermions, a high-frequency peak appears for
$n_{\uparrow}\geq0.5$, increasing in amplitude as $n_{\uparrow}$ grows
above the threshold while the doublon peak decreases, signaling the
transition of the dynamics to a regime dominated by the virtual
breaking of the pair (c.f. \ref{sec:high-frequency-peak}). Even for a
substantial value of $W$, $W/J=20$, the value of the oscillation
frequency is higher than in the fermionic case, signalling an
incomplete transition to a Tonks regime for the bosonic system.

% the effective area is clearly % smaller than in the fermionic case
for a range of densities, as % evidenced by an upshift in oscillation
frequency over the fermionic % case at equal density, and only becomes
comparable to the size of the % effective area for Fermions when
$n_{\uparrow}\approx0.9$.

On the other hand, when the boson-boson repulsion $W$ becomes
comparable to or smaller than the magnitude of the boson-impurity
attraction $|U|$, the numerics clearly show that higher-order bound
states between impurity and bath particles are formed in the ground
state. At $W/J=|U|/J=10$, both doublon and trion states (the impurity
binding to one or two bath particles on-site respectively) are
present, whereas, for $W/J=4$ doublon, trion and quatrion bound states
are occupied, with the trion state carrying the largest weight at any
bath density.  When quasi-momentum is applied to the impurity by the
kick, these higher order bound states perform oscillations, at
frequencies significantly lower than those for the doublons due to the
even higher effective mass. Interestingly, the damping we observe
becomes gradually smaller the smaller $W/J$ is, with the oscillations
at $W/J=4$ showing almost no decay at any value of $n_{\uparrow}$ in
the time domain over which we simulate. Whether this effect is due to
the partial ability of 1D superfluids to be protected against
excitations \cite{Kane:1992kx,Giamarchi:2003} -- which would be the
source of any damping of the bound state oscillations - is an
interesting question for further study.

\section{Discussion and conclusions}
\label{sec:disc_conc} The dynamics of the impurity moving on a 1D
lattice inside a fermionic bath, or a strongly repulsive bosonic one,
shows intriguingly complex dynamics, as can be seen by studying the
time-evolution of the doublon density. One of the characteristic
frequencies that appears corresponds to the motion of a free particle,
appearing in the limit of high filling of the bath and is easily
understood due to the increasingly uniform interaction energy the
impurity experiences on all sites.  Another feature of the dynamics,
namely the oscillations of a bound pair are also rather intuitive to
understand, allowing to draw an analogy to the molecule vs. polaron
question in three dimensional continuum systems.  The bound pair is
present in regime of large $|U|/J$ and low density, like the molecule
in the polaron vs.  molecule analogy. In the limit of small $|U|/J$ ,
we observed the dynamics of a free particle side-by-side pair breaking
of the correlated states of a polaron. This actually corresponds well
to the polaron ansatz \cite{Chevy:2006hv} which is a superposition of
a non-interacting Fermi sea (free particle) plus a contribution from
correlated particle - hole states. Thus, the crossover from a polaron
to a bound pair with increasing interaction is also taking place in
analogy to higher dimensional continuum systems.  In our system,
however, we have a feature that does not have any analogy in the
polaron vs.molecule crossover in 3D continuum, namely, the
high-frequency peak in the large $|U|=J$ regime, which becomes
dominant for bath fillings above 0.5, which is the result of a virtual
particle-hole creation process. This virtual exchange of paired and
bath particles can be read as a kind of internal dynamics of the
polaron. Observation of the dynamics predicted here should be feasible
in currently available ultracold gases systems, provided that the
temperature is below the energy scale of the oscillations which we
found to be of the order of $0.1 J$-$1 J$.

\begin{table}[!h] \centering
\begin{tabular}{|l|l|l|} \hline \hline $|U|$ range & Bath population &
Dynamics regime \\ \hline \hline Strong interaction & Large
$N_\uparrow$ & Free particle \\ \hline Strong interaction &
Intermediate $N_\uparrow$ & Bound pair + polaron internal dynamics \\
\hline Strong interaction & Small $N_\uparrow$ & Bound pair \\ \hline
Weak interaction & Large $N_\uparrow$ & Free particle \\ \hline Weak
interaction & Intermediate \& small $N_\uparrow$ & Free particle +
polaron \\ \hline \hline
\end{tabular}
\caption{Table summarising the different regimes identified for the
problem considered here.}
\label{tab:myfirsttable}
\end{table} \medskip {\it Acknowledgements} We thank J.\ Kajala for
useful discussions. This work was supported by the Academy of Finland
through its Centres of Excellence Programme (projects No. 251748,
No. 263347, No. 135000 and No.  141039), by ERC (Grant
No. 240362-Heattronics) and by the Swiss NSF under MaNEP and Division
II. Work in Pittsburgh is supported by NSF Grant
PHY-1148957. Computing resources were provided by CSC, the Finnish IT
Centre for Science.

\section{Appendix}
\subsection{Effective mass from the Bethe ansatz}\label{app:bethe} We
show here how to gain some insight into the problem through the string
hypothesis for the solution of the Lieb-Wu equations, whose solutions
describe (most of) the eigenvalues of the Hubbard Hamiltonian in one
dimension (Chapter 4 of \cite{Essler:2005uw}), in the limit of large
lattice lengths $L$.  For a fixed total number of particles $N$ and
number of down particles $M$, the patterns of which the solutions of
the Lieb-Wu equations are composed can be classified in three
different categories:
\begin{itemize}
\item \textbf{$k-\Lambda$} strings;
\item single real values of \textbf{$k_j$};
\item \textbf{$\Lambda$} strings.
\end{itemize} Every eigenstate of the Hubbard Hamiltonian can be
represented in terms of a particular configuration of strings,
containing $M_n$ $\Lambda$-strings, $M_n'$ $k-\Lambda$ strings of
length $n$ (in our case $n=1$), and $\mathcal{M}_e$ single $k_j$.
Here $M_n$, $M_n'$, $\mathcal{M}_e$ are related to the $N$ and $M$ by
\begin{align}
  \label{eq:6a} M=\sum_{n=1}^\infty n \left(M_n+M_n'\right) \\
N=\mathcal{M}_e+\sum_{n=1}^\infty 2n M_n'.
\end{align} In our case $M=1$ implies the existence of two classes of
string solutions
\begin{itemize}
\item[A)] $M_n=1$, $M_n'=0$, $\mathcal{M}_e=L$: this solution is
characterised by one $\Lambda$ string constituted by a single real
value, $N$ real $k_j$s and no $k-\Lambda$ strings.
\item[B)] $M_n=0$, $M_n'=1$, $\mathcal{M}_e=L-2$: in this case the
solution is characterised by $N-2$ real $k_j$s and one $k-\Lambda$
string, characterised by two (complex-valued) $k_{1,2}$s and one real
$\Lambda$, related by
 \begin{equation}
   \label{eq:14a} \sin(k_{1,2})=\Lambda\pm iU/4J .
 \end{equation}
\end{itemize}

In terms of the string parameters, energy and quasi-momentum are given
by
\begin{align} P&=\left[\sum_{j=1}^{N-2M'} k_j - \sum_{n=1}^\infty
\sum_{\alpha=1}^{M_n'}\left(2 \operatorname{Re} \arcsin\left(
{{\Lambda'}_\alpha} ^n + n i U/4J\right)-(n+1) \pi \right) \right]
\mod 2\pi \\
  \label{eq:7a} E&=-2 J \sum_{j=1}^{N-2M'} \cos\left(k_j\right) + 4
J\sum_{n=1}^{\infty}\sum_{\alpha=1}^{M'_n} \operatorname{Re} \sqrt{1-
\left( {{\Lambda'}_\alpha} + n i U/4J \right)^2}.
\end{align} For the two classes of solutions previously identified,
see A and B above, take the following form
\begin{itemize}
\item[A)]
  \begin{align}
    \label{eq:8a} P&= \sum_{j=1}^{N} k_j \\ E&= -2 J\sum_{j=1}^N
\cos\left(k_j\right)+ const.
  \end{align}
\item[B)]
  \begin{align}
    \label{eq:9a} P&= \sum_{j=1}^{N-1} k_j -2 \operatorname{Re} \left[
\arcsin\left( \Lambda + i U/4J\right) \right] \\ E&= -2 J
\sum_j^{N-1}\cos\left(k_j \right) + 4 J \operatorname{Re}
\sqrt{1-\left( \Lambda + i U/4J \right)^2} + const.
  \end{align}
\end{itemize} The solution containing the $k-\Lambda$ string can be
written in a more transparent form as
\begin{align}
  \label{eq:10a} P&= \sum_{j=1}^{N-1} k_j + 2 q \\ E&= -2 J
\sum_j^{N-1} \cos\left(k_j \right) - 4 J \cosh \xi \cos \left( q
\right) + const.
\end{align} where $q=\operatorname{Re} \left[ k_1
\right]=\operatorname{Re} \left[ k_2 \right]$ and
$\xi=\operatorname{Im} \left[k_1\right]=-\operatorname{Im}
\left[k_2\right]$, with $k_1$ and $k_2$ belonging to the $k-\Lambda$
string.  For the $k-\Lambda$ string it is possible to prove (see
\cite{Essler:2005uw} p. 134) that
\begin{equation}
  \label{eq:11a} \cosh \xi = \sqrt{1+\frac{U^2}{16 J^2
\cos^2\left(q\right)}}.
\end{equation} From the form of the wavefunction associated with the
$k-\Lambda$ string, $\xi$ describes the (exponential) spatial decay of
the pair, and thus $\xi^{-1}$ can be interpreted as the size of the
pair.
\begin{figure}%[!ht]
  \includegraphics[width=0.5\textwidth]{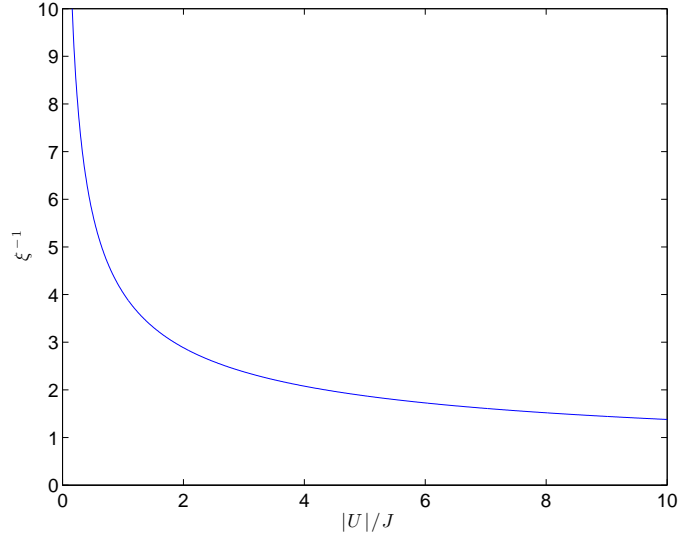}
  \caption{Pair size ($\xi^{-1}$) as a function of $U$.}
 \label{fig:psize}
\end{figure}

In the strongly interacting limit we have
\begin{equation}
  \label{eq:12a} -4 J \cosh \xi \cos q \to U -4 J^2 / \left| U \right|
-4 J^2 / \left| U \right| \cos \left(2 q \right)
+O\left(\frac{1}{U^2}\right)
\end{equation} which is coherent with the strong coupling calculation
leading to the Heisenberg Hamiltonian for the Hubbard Hamiltonian in
the strong coupling limit (modulo a $U \to -U$ mapping).

The spectrum is thus constituted by a lower band ($k-\Lambda$+ single
$k$s solutions) and a higher band (single $k$s and single $\Lambda$
solutions).  Intuitively, the former correspond to a pair in an
unpaired background Fermi sea, while the latter to scattering
states. The down particle which is kicked in our simulations, with a
view to the collective nature of the excitations in 1D systems, can be
considered as constituted by both kinds of elementary excitations,
appropriately weighted by the presence of the trapping potential.

We now aim at describing the dynamics that we observe numerically
through the evaluation of the effective mass for the pairs
\begin{align}
  \label{eq:13} m_{\rm doublon}&=\left. \left(\frac{\partial^2 E
}{\partial \kappa^2}\right)^{-1}\right|_{\kappa=0}\\ &=
\frac{1}{J}\sqrt{1+\frac{U^2}{16 J^2}},
\end{align} where $E$ is defined by Eq. \eqref{eq:10a}.

\subsection{The frequency of the non-interacting particle in the
combined harmonic trap and lattice potential}

The idea is to compare the oscillation frequency from the numerical
data to the one given by the non-interacting particle in a combined
lattice and harmonic trap potential. As a reminder, the formula for
the centre of mass (COM) position was given by
\begin{equation}
  \label{eq:COMa} \langle x \rangle = \delta \exp \left[ \left(
\frac{\delta}{a_{\rm ho}}\right)^2 \sin^2\left(V/8 t\right)\right]
\cos\left[V\left(\sqrt{\frac{2}{V m_{\rm eff}}}-\frac{1}{4}\right)t-
\frac{\delta^2}{2a_{\rm ho}^2}\sin\left(\frac{V
t}{4}\right)+\frac{\pi}{2}\right].
\end{equation} Neglecting the exponential prefactor time dependence,
we can write the centre-of-mass oscillation frequency as
\begin{equation}
  \label{eq:14} \omega_{COM}=\frac{d \Omega}{d t} ,
\end{equation} where
$$
\Omega=V\left(\sqrt{\frac{2}{V m_{\rm eff}}}-\frac{1}{4}\right)t-
\left( \frac{\delta}{a_{\rm ho}}\right)^2\sin\left(\frac{V
t}{4}\right)+\frac{\pi}{2}
$$
with $\delta=\left[\frac{\left(1-\cos k \right)}{V m_{\rm
eff}}-\frac{1}{16}-\frac{ \sqrt{V/J}}{256}\right]^{1/2}$, leading to
\begin{equation}
  \label{eq:1} \omega_{COM}= V\left[\left(\sqrt{\frac{2}{m_{\rm eff}
V}}-\frac{1}{4}\right)-\frac{\sqrt{V/J}}{4} \left(\frac{1-\cos
k}{m_{\rm eff} V}-\frac{1}{16}-\frac{\sqrt{V/J}}{256}\right)\right].
\end{equation} The comparison between $\omega_{COM}$ and the numerical
data is obtained by performing a discrete Fourier transform of
$\langle n_{\uparrow}n_{\downarrow}\rangle(t)$ for different values of
$U$ and $N_{\uparrow}$. For $N_{\uparrow}=N_{\downarrow}=1$, i.e. no
bath, the agreement is perfect: higher interaction energies correspond
to lower values of the oscillation frequency, in agreement with the
increase of the effective mass.

\end{document}